\documentclass[graybox]{svmult}

\usepackage{mathptmx}       
\usepackage{helvet}         
\usepackage{courier}        
\usepackage{type1cm}        
\usepackage{makeidx}         
\usepackage{graphicx}        
\usepackage{multicol}        
 \usepackage[stable]{footmisc}

\usepackage{mathtools}
\usepackage{amsfonts} 
\usepackage{amsmath}
\usepackage{bbold}

\usepackage{stix}
\usepackage{amsmath,mleftright,mathtools}
\DeclareMathAlphabet\mathbfcal{OMS}{cmsy}{b}{n}

\usepackage{dsfont}

\def\n{^{(n)}}
\def\pr{^{\prime}}
\def\Sigmab{\boldsymbol\Sigma}

\makeindex             

\begin{document}

\title*{Dynamic Factor Models: a Genealogy}
\author{Matteo Barigozzi and Marc Hallin}
\institute{Matteo Barigozzi \at Dipartimento di Scienze Economiche \at Alma Mater Studiorum,  Universit\`a di Bologna, Italy\\ \email{matteo.barigozzi@unibo.it}
 \\  $\,$ \\ 
Marc Halllin  \at ECARES and Department of Mathematics \at Universit\' e libre de Bruxelles, 
Belgium\\ \email{mhallin@ulb.ac.be}}
%
\maketitle


\abstract{Dynamic factor models have been developed out of the need of analyzing and forecasting time series in increasingly high dimensions. While mathematical statisticians faced with inference problems in high-dimensional observation spaces were focusing on the so-called {\it spiked-model-asymptotics}, econometricians adopted an entirely and considerably more effective asymptotic approach, rooted in the factor models originally considered in psychometrics. The so-called {\it dynamic factor model} methods, in two decades, has grown into  a wide and successful body of techniques that are widely used in  central banks, financial institutions, economic and statistical institutes. The objective of this chapter is not an extensive  survey   of the topic but a sketch of its historical growth, with emphasis on the various assumptions and interpretations, and a family tree of its main  variants. 
}

%

\section{Factor models and the analysis of high-dimensional time series}
With the fast-pace development of computing facilities, high-dimensional datasets are increasingly available, posing a genuine challenge to statisticians and econometricians.  Faced with this situation and the need to analyze such datasets, new asymptotic scenarios and methods had to be developed. Mathematical statisticians mostly focused on the so-called spiked models (see, for instance, Johnstone~(2001), Onatski et al.~(2013, 2014)), which leads to beautiful mathematical results such as the  phase transition phenomenon, the Mar\v cenko-Pastur, and the Tracy-Widom laws but also, due to the ``fixed-sized needle in a growing haystack'' nature of their asymptotic scenario (see Hallin~(2023) for details), somewhat limited practical consequences. More realistic  asymptotics, of the ``growing needle in a growing haystack''  type, simultaneously were adopted by econometricians, which led to the by far more  successful  dynamic factor methods. 

The objective of this paper is a historical account of the emergence of this dynamic factor  approach to the analysis of high-dimensional time series, with emphasis on the assumptions, their interpretation, and the ``genealogic''  aspects of its development. The bibliography is unavoidably limited, and even sketchy; it unevitably but involuntarily overlooks a number of relevant contributions and  reflects personal, hence biased, views for which we apologize.

\begin{figure}[htbp]
\centering
\vskip 1cm
\rotatebox{90}
{\includegraphics[width=1.6\textwidth, trim=0mm 0mm 0mm 0mm, clip]
{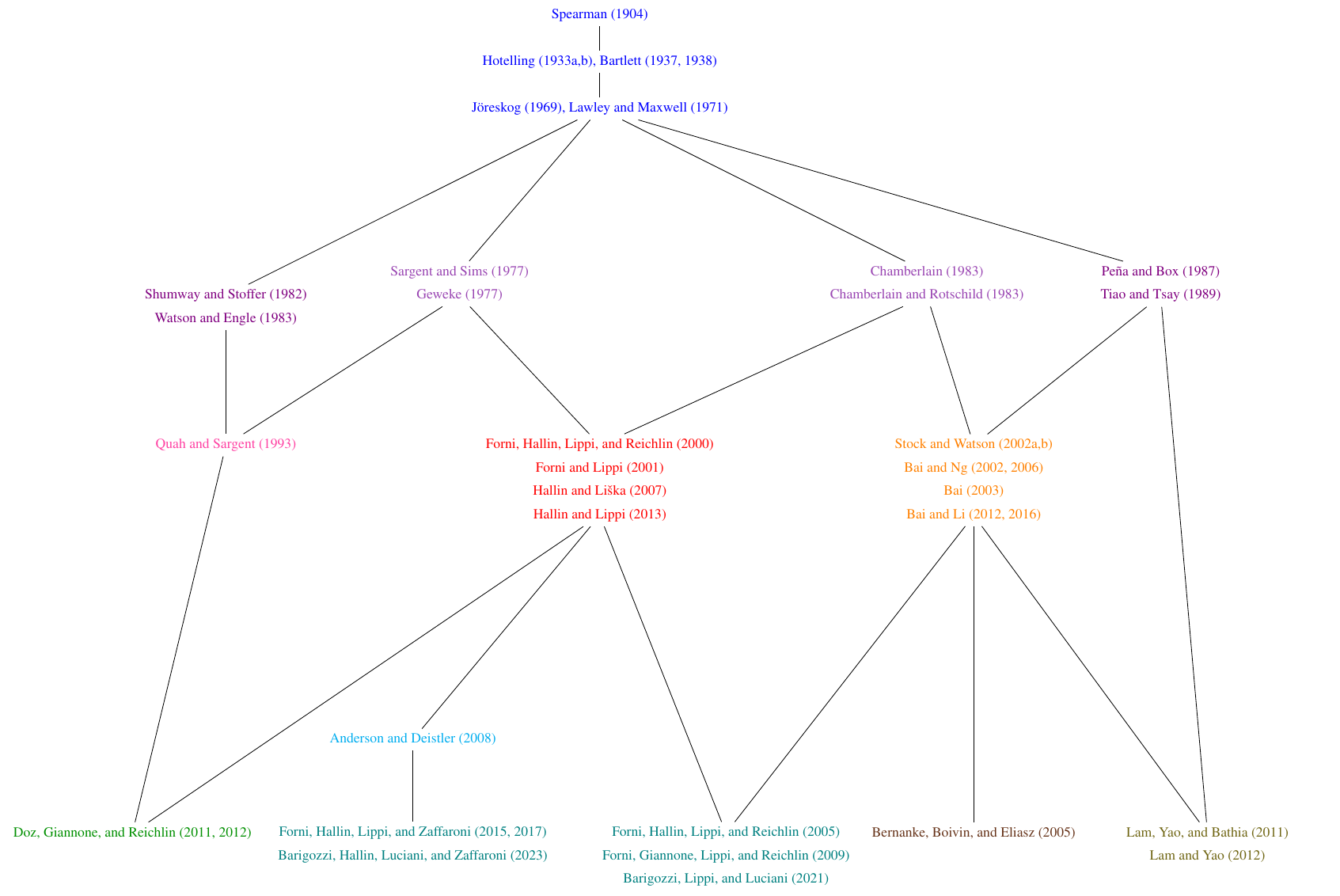}
}
\end{figure}

\section{Remote ancestry: Spearman (1904) and the psychometric roots of factor models}\label{Sec2}
The family tree of factor models is rooted into early-century psychometrics, and it is usually admitted that the concept first appears more than a century ago in  Spearman~(1904). Spearman proposes factor analysis in order to account for the dependencies between  several variables related with cognitive abilities   measured on  given individuals. The result was a two-factor theory in which  cognitive performance was explained by two unobservable  ``factors'':   {\it general ability}  and a second one  which later on was dropped as non-significant. The concept of  IQ or intelligence quotient  usually refers to this general mental ability factor.

The objective of factor models, thus, is to account for cross-sectional depen\-dencies---more specifically, cross-sectional covariances or correlations. Spearman's exposition does not match the mathematical standards of present-day psychometrics or statistics, and more precise mathematical descriptions of factor models only came somewhat later. Classical references are, among others, Hotelling~(1933a,b), Bartlett~(1937, 1938), J\" oreskog (1969); see also Lawley and Maxwell~(1971) for a classical textbook exposition, and J\" oreskog~(2007) for a historical account. 

In modern language, a factor model with $r$ factors (it is expected that~$r\ll n$,  where~$n$ is the number of components of the observed variable $X$) is a statistical model characterized by an equation of the form 
\begin{equation}\label{1}
{\bf X}_t =\boldsymbol{\chi}_t + \boldsymbol{\xi}_t \coloneqq  {\bf B}{\bf f}_t + \boldsymbol{\epsilon}_t, \quad t= 1,\ldots T,
\end{equation}
where 
\begin{enumerate}
\item[(a)] ${\bf X}_1,\ldots,{\bf X}_T$ is an i.i.d.\   sample of $n$-dimensional observations ${\bf X}_t=(X_{1t},\ldots,X_{nt})^\prime$ with (for ease of exposition and  without any  loss of generality) zero-mean, strictly positive variance, and finite second-order moments;
\item[(b)] ${\bf B }= (B_{ik})_{1\leq i\leq n ,\: 1\leq k\leq r}$ is an $n\times r$   matrix of scalar {\it loadings};\vspace{1mm}
\item[(c)] ${\bf f}_t=(f_{1t},\ldots ,{f}_{rt})^\prime$ is an i.i.d.\  process of latent (unobservable) $r$-dimensional variables, the {\it (common) factors} with zero-mean and unit variance;
\item[(d)]  $\boldsymbol{\epsilon}_t=(\epsilon_{1t},\ldots ,\epsilon_{nt})^\prime$, called the {\it idiosyncratic component}, is an i.i.d. process of $n$-dimensional zero-mean variables with finite diagonal covariance matrix.
\end{enumerate}
This equation decomposes the observation $X_{it}$ into an unosbservable {\it common component} ${\chi}_{it} $ and an unobservable  {\it idiosyncratic component} ${\xi}_{it}=\epsilon_{it}$. In order to identify this decomposition, 
it is assumed, moreover, that  the following  orthogonality conditions hold: 
\begin{enumerate}
\item[(i)] ${\rm E}[{f}_{kt}{f}_{\ell t}]=0$, \  $1\leq k\neq\ell\leq r,\  \ 1\leq t \leq T$;
\item[(ii)] ${\rm E}[{f}_{kt}{\epsilon}_{j t^\prime}]=0$, $1\leq k \leq r,\  1\leq j \leq n, \ 1\leq t,\ t^\prime \leq T$;
\end{enumerate}

In this statistical model, the space spanned by the loadings (that is, by the columns of $\bf B$)  is identified, while the loadings $\bf B$ themselves and the factors  ${\bf f}_t$ and the  loadings $\bf B$  are only identified up to pre- and post-multiplication by an arbitrary $r\times r$ orthogonal matrix   $\bf O$ and its inverse~${\bf O}^\prime$: indeed,  $\bf B{\bf f}_t = \bf B{\bf O}^\prime \bf O{\bf f}_t$ for any such matrix. 
 This indeterminacy, however, should not be interpreted as a weakness---quite on the contrary, it provides a quite precious flexibility in the choice of interpretable factors. 

Usual estimation methods are based on Gaussian maximum likelihood, and consistency (up to an orthogonal transformation) of the estimated loadings $\hat{\bf B}^{(T)}$ is achieved for fixed $n$ as~$T\to\infty$; we refer to Anderson and Rubin~(1956) and Amemiya, et al.~(1987) for a comprehensive coverage of the subject. 

This consistency of $\hat{\bf B}^{(T)}$ as $T\to\infty$, however, does not allow for a   consistent recovery of the  factors ${\bf f}_t$.  The latter have to be retrieved,  for each   $t$, as the linear projections $\hat{\bf f}_t^{(T)}\coloneqq (\hat{\bf B}^{(T)\prime}\hat{\bf B}^{(T)})^{-1}\hat{\bf B}^{(T)\prime}{\bf X}_t$ of ${\bf X}_t$ onto the estimated loadings; now, 
\begin{align*}
(\hat{\bf B}^{(T)\prime}\hat{\bf B}^{(T)})^{-1}\hat{\bf B}^{(T)\prime}{\bf X}_t  
&= ({\bf B}^{\prime}{\bf B})^{-1}{\bf B}^{\prime}{\bf Bf}_t + ({\bf B}^{\prime}{\bf B})^{-1}{\bf B}^{\prime}{\boldsymbol \epsilon}_t + o_{\rm P}(1)\\
&= {\bf f}_t + ({\bf B}^{\prime}{\bf B})^{-1}{\bf B}^{\prime}{\boldsymbol \epsilon}_t + o_{\rm P}(1) 
\end{align*}
as $T\to\infty$, where $({\bf B}^{\prime}{\bf B})^{-1}{\bf B}^{\prime}{\boldsymbol \epsilon}_t $, which does not depend on $T$, is not $ o_{\rm P}(1)$ as  $T\to~\!\infty$.  On the other hand, in view of Assumption~(d), if we let the dimension $n$ of ${\bf X}_t$ tend to infinity with, for instance,  bounded  (uniformly in $n$) diagonal   idiosyncratic covariance matrix elements, $({\bf B}^{\prime}{\bf B})^{-1}{\bf B}^{\prime}{\boldsymbol \epsilon}_t $ is~$o_{\rm P}(1)$ as $n\to\infty$. This is the first sign that high-dimensional asymptotics are  ideal in the context of factor models---the ``blessing of dimensionality'' phenomenon that will play a major  behind-the-scenes role in the subsequent developments of factor models.

While the factors, in \eqref{1}, are accounting for all cross-sectional covariances, they do not necessarily account for  variances, since idiosyncratic variances can be quite high. Dropping the idiosyncratic component in \eqref{1} sometimes is interpreted as a dimension reduction technique, which may be misleading and quite dangerous in the presence of large idiosyncratic variances. This is in sharp contrast with Principal Component Analysis (PCA, introduced by Pearson~(1901), three years before Spearman's  first paper on factor models), where  dropping the last eigenvalues and eigenvectors has a small impact on the total variance of the observation. 

Note also that Equation~(1), along with~conditions (a)--(d) and~(i)--(ii), constitutes a {\it statistical model}---that is,   imposes  on~${\bf X}_t$ restrictions that the distribution of a typical random vector~${\bf X}_t$ does not satisfy. Before performing a factor analysis on ${\bf X}_t$, thus, one should be cautious and check whether~${\bf X}_t$ satisfies the model assumptions.  When it does, \eqref {1} may or may not be interpreted as describing a data-generating process of the ``signal plus noise'' type, with the idiosyncratic component playing the role of noise. This, again, is in sharp contrast with PCA, which is ``model-free'': provided that its second-order moment are finite, indeed,  {\it any} ${\bf X}_t$ admits a decomposition into principal components.  Tipping and Bishop (1999) showed that if we also assume homoskedasticity of idiosyncratic components then PCA estimates are equivalent to Gaussian maximum likelihood ones. The use of PCA to estimate factor models was first  proposed by Hotelling (1933a,b), and then almost forgotten for quite some time, see Section 4. \medskip

As we shall see, all factor models are based, as Spearman's original one   in \eqref{1}, on a decomposition of the observation~$X_{it}$ into the sum~$\chi_{it}+~\!\xi_{it}$ of a common and an idiosyncratic component; they  only differ by the various conditions ((a)--(d) \linebreak and~(i)--(ii)) imposed on this decomposition, which characterize various notions of ``commonness'' and ``idiosyncrasy.''

\section{The pathbreaking   generation: Geweke (1977), Sargent and Sims~(1977), Chamberlain~(1983), Chamberlain and Rothschild~(1983) }

In the late 1970s, it appeared that traditional econometric time series models---typically, VAR and VARMA models---were defeated by the curse of dimen\-siona\-lity~and the increasingly high dimensions of the econometric  series to be analyzed and predicted. A simple~$n$-dimensional VAR(1) model, for instance, involves no less than~$n(3n+1)/2$ parameters ($n^2$ autoregressive coefficients, $n$ innovation variances, and $n(n-1)/2$ innovation covariances).  This means 610 parameters for $n=20$, 15,050 for $n=100$, and~375,250 for $n=1000$! A solution had to be found, able  to deal with both the high-dimensional aspect of the data and   their time series nature. The first steps towards that solution,  based on serial extensions of the traditional factor model of Section~\ref{Sec2},  were taken in four pathbreaking papers: Geweke (1977), Sargent and Sims~(1977), Chamberlain (1983), and Chamberlain and Rothschild~(1983).  The publication dates (running between~1977 and 1983) are mainly due to editorial \linebreak hazards and do not reflect any significant precedence.\

\subsection{Geweke (1977) and dynamic loadings}
Geweke (1977), shortly followed by Sargent and Sims (1977),  first understood that, if factor models were to be used in standard econometric problems, some of the conditions imposed in Section~\ref{Sec2} were to be relaxed. To begin with, the time-series nature of econometric data cannot be ignored: the  i.i.d-ness assumptions  in (a) and~(c)   cannot be maintained. Second---and this is an extremely innovative idea---the value~$f_{kt}$ of a factor at time $t$ may be loaded by the observation with some lag: instead of contemporaneous loadings via a loading matrix ${\bf B}$ (call them {\it static loadings)}, Geweke considers {\it dynamic loadings} via {\it loading filters}  ${\bf B}(L)$ where ${\bf B}(L)=\sum_{\nu=0}^\infty {\bf B}_\nu L^\nu$ 
 is an $n\times r$ 
    matrix of one-sided filters with square-summable entries.
    Here $L$, as usual, stands for the {\it lag operator}. 

Geweke's is a statistical model characterized by an equation of the form 
\begin{equation}\label{2}
{\bf X}_t =\boldsymbol{\chi}_t + \boldsymbol{\xi}_t \coloneqq {\bf B}(L){\bf f}_t + \boldsymbol{\epsilon}_t = \sum_{\nu = 0}^\infty {\bf B}_{\nu}{\bf f}_{k, t-\nu} + \boldsymbol{\epsilon}_t , \quad t= 1,\ldots T,
\end{equation}
where 
\begin{enumerate}
\item[(a$^\prime$)] ${\bf X}_1,\ldots,{\bf X}_T$ is the finite realization an observed $n$-dimensional  second-order stationary process ${\bf X}=(X_{1t},\ldots,X_{nt})^\prime$ with (for ease of exposition and  without any real loss of generality) zero-mean, strictly positive variance,  and finite second-order moments;
\item[(b$^\prime$)] ${\bf B}(L) \coloneqq \left(\sum_{\nu =0}^\infty B_{ik\nu}L^\nu\right)_{{1\le i\le n,\: 1\le k\le r}}$\vspace{1mm}   is an $n\times r$   matrix of  {\it loading filters} with square-summable coefficients for any   $i$ and $k$;
\item[(c$^\prime$)] ${\bf f}_t=(f_{1t},\ldots ,{f}_{rt})^\prime$ is a second-order stationary  latent   $r$-dimensional process of {\it factors} with ${\rm E}[f_{kt}]=0$ and ${\rm E}[f^2_{kt}]=1$ for $k=1,\ldots,r$ and~$t\in{\mathbb Z}$; 
\item[(d$^\prime$)] $\boldsymbol{\epsilon}_t=(\epsilon_{1t},\ldots ,\epsilon_{nt})^\prime$, called the {\it idiosyncratic component}, is an $n$-dimensional second-order stationary white noise process with finite diagonal covariance matrix. 
\end{enumerate}
It is assumed, moreover, that  the following  orthogonality conditions hold: 
\begin{enumerate}
\item[(i$^\prime$)] ${\rm E}[{f}_{kt}{f}_{\ell t^\prime}]=0$, \  $1\leq k\neq\ell\leq r,\  \ 1\leq t, t^\prime \leq T$;
\item[(ii)] ${\rm E}[{f}_{kt}{\epsilon}_{j t^\prime}]=0$, $1\leq k \leq r,\  1\leq j \leq n, \ 1\leq t,\ t^\prime \leq T$.
\end{enumerate}
Equation~\eqref{2} is an extension of the classical equation~\eqref{1}; Assumptions (a$^\prime$)--(d$^\prime$), clearly, are relaxations of (a)--(d); condition (i$^\prime$) is reinforcing condition~(i) by requiring orthogonality of the factors at all leads and lags; condition~(ii) remains unchanged. 
These conditions define an {\it exact  dynamic}  factor model.

Sargent and Sims~(1977), under the name {\it unobservable index model}, and Geweke and Singleton (1981) provide an equivalent frequency-domain description (which we do not reproduce here) under the additional assumption, of course,  of the existence of a spectrum for~${\bf f}_t$ and~${\boldsymbol\epsilon}_t$, hence for ${\bf X}_t$.  Thanks to the assumption (d$^\prime$) that the idiosyncratic processes are mutually orthogonal white noises (their lagged cross-covariances  all are zero), the  exact dynamic factor model is fully identified (again, up to an orthogonal transformation of the factors) for fixed $n$; {see Geweke and Singleton~(1981)}. 

{Estimation is typically performed by means of spectral Gaussian maximum likelihood, see Sargent and Sims~(1977) and recent work by Fiorentini et al. (2018). As in the classical model , consistency  as $T\to\infty$ with $n$ fixed  is possible only for the estimated loadings, not for the factors themselves.}

\subsection{Chamberlain (1983) and the approximate factor model}

In Geweke's  approach, the dimension $n$ of the observation is fixed. His  assumption~(d$^\prime$)  of  mutually orthogonal white noise idiosyncratic components, moreover, is quite unlikely to hold in practical situations---all the more so when~$n$ is large: it requires, indeed, that {\it all} cross-covariances be accounted for by the $r$ factors. On the other hand, removing this condition (d$^\prime$) results in an unidentifiable model where the factor-driven and the idiosyncratic components ${\boldsymbol\chi}_t$ and ${\boldsymbol\xi}_t$ cannot be sepa\-rated as $T\to\infty$. Finally, the high-dimensional nature of the observations, in a \linebreak fixed-$n$ approach, is not fully taken into account and the ``blessing of dimensionality,'' moreover, is not exploited. This motivated Chamberlain~(1983) and Chamberlain and Rothschild~(1983) to consider an asymptotic scheme under which both $n$ and $T$ tend to infinity---a double-asymptotics approach that has, since then, become standard in high-dimensional statistics. The~$(n, T(n))$ path along which this double-asymptotics scheme is achieved, i.e., the relative magnitude of $n$ and $T$, in most  identification and consistency results, plays no asymptotic role, though: in particular,~$n$ can be larger than $T$.

The model developed by Chamberlain (1983) and by Chamberlain and Rothschild~(1983) comes back to the classical  equation 
\begin{equation}\label{3}
{\bf X}_t^{(n)} =\boldsymbol{\chi}_t^{(n)} + \boldsymbol{\xi}_t^{(n)} \coloneqq {\bf B}^{(n)}{\bf f}_t + \boldsymbol{\epsilon}_t^{(n)} 
\quad t= 1,\ldots T,\ \ n\in {\mathbb N}. 
\end{equation}

This equation   coincides with the traditional static-loading factor model of Equation \eqref{1},  with an important difference, though:  superscripts\! $^{(n)}$ have been added \linebreak to~${\bf X}_t$, ${\boldsymbol\chi}_t$, ${\boldsymbol\xi}_t$, ${\bf B}$,  and ${\boldsymbol \epsilon}_t$ in order to emphasize the fact that they are $n$-dimensional, with $n$  tending to infinity just as $T$ does. Chamberlain's  assumptions, moreover, are quite mild: 
  while (a$^\prime$) and (c$^\prime$) are borrowed from Geweke and~(b) from the traditional model,    Chamberlain relaxes Geweke's strong diagonality assumption~(d$^{\prime}$) into 
\begin{enumerate}
\item[(d$^{\prime\!\prime}$)] 
$\boldsymbol{\epsilon}_t^{(n)}=(\epsilon_{1t},\ldots ,\epsilon_{nt})^\prime$, called the {\it idiosyncratic component}, is an $n$-dimensional second-order stationary process with finite covariance matrix ${\boldsymbol\Sigma}^{(n)}_{\boldsymbol\epsilon}$.
%
%
\end{enumerate}
This covariance matrix  ${\boldsymbol\Sigma}^{(n)}_{\boldsymbol\epsilon}$ needs not be diagonal.

As for the orthogonality conditions, Geweke's conditions (i$^\prime$) and (ii$^\prime$) are kept unchanged, but we add two new conditions: 
\begin{enumerate}
\item[(iii)] denoting by $\lambda ^{(n)}_{\boldsymbol{\chi};1}\geq\lambda ^{(n)}_{\boldsymbol{\chi};2}\geq\ldots\geq \lambda ^{(n)}_{\boldsymbol{\chi};r}$ the eigenvalues of the $n\times n$ rank $r$ covariance matrix~${\boldsymbol\Sigma}^{(n)}_{\boldsymbol{\chi}}$ of $\boldsymbol{\chi}_t^{(n)}$,  $\lim_{n\to\infty} \lambda ^{(n)}_{\boldsymbol{\chi};r}=\infty$;
%
%
\item[(iv)] denoting by $\lambda ^{(n)}_{\boldsymbol{\epsilon};1}\geq\lambda ^{(n)}_{\boldsymbol{\epsilon};2}\geq\ldots\geq \lambda ^{(n)}_{\boldsymbol{\epsilon};n}$ the eigenvalues of the $n\times n$ full-rank covariance matrix~${\boldsymbol\Sigma}^{(n)}_{\boldsymbol{\epsilon}}$ of $\boldsymbol{\epsilon}_t^{(n)}$,  $\lim_{n\to\infty}\lambda ^{(n)}_{{\boldsymbol\epsilon};1} <\infty$. 
\end{enumerate}
These assumptions define an \textit{approximate static factor model}. A similar setting was independently proposed by Connor and Korajczyk (1986).\medskip 

A price is to be paid, however, for relaxing the diagonality assumption on ${\boldsymbol\Sigma}^{(n)}_{\boldsymbol\epsilon}$: the model is no longer identifiable for fixed $n$ and hence, still for fixed $n$, is no longer consistently estimable as $T\to\infty$. If, however, the cross-sectional dimension $n$ tends to infinity, identifiability resurfaces under asymptotic form. Indeed, if ${\bf X}_t^{(n)}$ satisfies  Equation~\eqref{3} and the assumptions (a$^\prime$), (b), (c$^\prime$), (d$^{\prime\prime}$), (i)$^\prime$, (ii$^{\prime\prime}$), (iii), and (iv) of the approximate static factor model,  it follows from   (iii) and (iv) and a straightforward application of Weyl's inequality  that  there exists  a finite~$r\in{\mathbb N}$ independent of $n$ such that, denoting by $\lambda ^{(n)}_{{\bf X};1}\geq\lambda ^{(n)}_{{\bf X};2}\geq\ldots\geq \lambda ^{(n)}_{{\bf X};n}$ the eigenvalues of the $n\times n$  covariance matrix~${\boldsymbol\Sigma}^{(n)}_{\bf{X}}$ of ${\bf{X}}_t^{(n)}$,   
\[ 
\lim_{n\to\infty}\lambda ^{(n)}_{{\bf X};r} =\infty\quad \text{ and }\quad \lim_{n\to\infty}\lambda ^{(n)}_{{\bf X};r+1}  <\infty .
\]

Usual estimation approaches are PCA and (pseudo-)Gaussian maximum likelihood. Since we now assume that both $n$ and $T$ tend to infinity, we can consistently estimate both the loadings and the factors (up to orthogonal transformations, as usual). Hence, it is possible to consistently estimate also the common components:  letting~$n\to\infty$ wards off the curse of dimensionality problem--- the previously mentioned    {\it ``blessing of dimensionality''} phenomenon.

\section{Laying the  modern foundations: 
Stock and Watson~(2002), Bai~(2003), Forni, Hallin, Lippi, and Reichlin (2000)}\label{sec3}

%
The powerful ideas of Geweke, Sargent and Sims, Chamberlain, and Rothschild were not fully exploited until, in the early 2000's,   they were picked up, almost simultaneously,  by three groups of econometricians working independently of each other. While the models developed by Stock and Watson~(2002a,b) and Bai~(2003) belong to the lineage of Chamberlain and Rothschild, with double asymptotics and static loadings, the model proposed by Forni et al.~(2000) and Forni and Lippi~(2001) is combining, under the name \textit{Generalized} or \textit{General  Dynamic Factor Model} the attractive features of Geweke~(1977) (dynamic loadings via filters rather than matrices) with those of Chamberlain~(1983) (an approximate factor model where both $n$ and~$T$ tend to infinity).  The publication dates (running between~2000 and 2003) are due to refereeing hazards and do not reflect any significant precedence.

\subsection{Static loadings and the  approximate factor model: Stock and Watson~(2002), Bai (2003), and some others}  

The factor model considered by Stock and Watson~(2002a,b) and Bai~(2003) (and Bai and Ng~(2002) in their paper on the identification of the number of factors: see Section~\ref{SecIdent}) is essentially the same approximate factor model as in Chamberlain~(1983) and Chamberlain and Rothschild~(1983)---with some minor variations in the assumptions.
These papers provide a rigorous treatment of the asymptotic properties of the PCA-based estimators of the model, and show, as expected, that if both $n$ and $T$ tend to infinity,  consistency (up to orthogonal transformations, as usual) is achieved for the loadings and the factors.  Typically, once factors are extracted via PCA from an~$n$-dimensional ($n$ large) time series ${\bf X}_t$, they are used in a second step to predict a given target variables. This approach, in general, offers sizeable improvements over univariate or small-$n$d forecasting models, see, e.g.,  Stock and Watson~(2002a,b) for  empirical evidence. 

These papers had a major impact on the econometric literature and largely contributed to the dissemination and the development of contemporary factor model methods.  Further notable developments along the same lines are the study of {\it factor augmented prediction} (Bai and Ng~2006) and the high-dimensional extension of classical results on pseudo-Gaussian maximum likelihood estimation (Bai and\linebreak  Li~2012, 2016).

Predating Stock and Watson~(2002a,b) and Bai  (2003), one also should mention two groups of earlier contributions which considered extensions of the exact factor model apt to capture specific aspects of the observed time series: 
\begin{enumerate}
\item[(I)] Engle and Watson (1981), Shumway and Stoffer (1982), Watson and Engle (1983), and Quah and Sargent (1993)  adopted  a ``state-space approach''' where a dynamic equation for the factors, e.g., a VAR specification, is added to the static factor model, specifying a parametric structure for the   factors'  autocorrelations;
\item[(II)] Pe\~{n}a and Box (1987) and Tiao and Tsay (1989)  revisited the exact static factor model in a time series context, thus assuming the idiosyncratic components to be a second-order stationary white noise process.
\end{enumerate}

{Approach (I) was extended to the high-dimensional setting $n\to\infty$ by Doz et al.~(2011, 2012) who considered the use of the Kalman filter combined with Gaussian maximum likelihood estimation via the Expectation Maximization algorithm. This approach is among of the most frequently used in macroeconomic policy analysis; it is employed for now-casting (Giannone et al. (2008)) and for building indicators of economic activity (Barigozzi and Luciani (2021)). See also Poncela et al. (2021), for a survey.}

{Approach (II) was extended to the high-dimensional setting $n\to\infty$ by Lam et al., (2011) and Lam and Yao (2012) who still consider  principal-component-based estimation but based on a sum of autocovariances---under an assumption of white noise idiosyncratic components which, however,  is unlikely to hold in practice.} 


\subsection{Dynamic  loadings and  the  { General Dynamic Factor Model}: Forni, Lippi, Hallin, and Reichlin (2000)}

The General or {\it Generalized} Factor Model (henceforth GDFM)  was proposed by Forni et al.~(2000) and Forni and Lippi~(2001). It is combining the dynamic loadings ideas of Geweke (1977) and Sargent and Sims (1977) with the double asymptotics ($n,T\to\infty$) of Chamberlain (1983) and Chamberlain and Rothschild (1983). Dynamic loadings allow for capturing the lagged impacts of the common factors driving the common component 


%

The following presentation is inspired from the time-domain exposition of Hallin and Lippi~(2013), which avoids the spectral-domain approach originally used by  Forni and Lippi~(2001) to derive the of the GDFM. Its spirit  also slightly differs from that of Lippi et al.~(2023). 

In this approach the observation, an  $n\times T$ panel,  is the finite  realization,\linebreak  for~$1\leq i\leq n$ and $1\leq t\leq T$ ($n$ and $T$ large), of 
 a double-indexed  stochastic process~$\mathbf{X}:=\{X_{it} \vert i\in\mathbb{N} ,  \  t\in\mathbb{Z}
\}
$,
that is, a collection of $n$   observed time series  of length~$T$,  related to $n$ individuals or ``cross-sectional items"  or, 
equivalently, one single  time series in dimension $n$.  We denote by~${\bf X}\n_t$ the $n$-dimensional vector $(X\n_{1t},\ldots , X\n_{nt})\pr$, 
  by ${\bf X}_t$ the fixed-$t$ collection~$\{X_{it}\vert i\in\mathbb{N}\}$,  and by~${\bf X}\n$ the $n$-dimensional process $\{X_{it} \vert 1\le i\le n ,  \  t\in\mathbb{Z}
\}$. It is assumed  throughout that 
 $\mathbf{X}$ is second-order time-stationary, i.e.,  for all  values of~$i$, $i\pr$, $i^{\prime\prime}$, $t$,  and~$k$,  the variances~Var$(X_{i t})$ and covariances   Cov$(X_{i\pr t}X_{i^{\prime\prime}, t-k})$   exist, are finite, and do not depend on $t$. For simplicity, we also assume that all~$X_{it}$'s have zero-mean centered and, in order to avoid trivialities, are non-degenerate: $\mathrm{E}[X_{it}]=0$ and $0<\mathrm{E}[X^2_{it}]<\infty $ for all~$i\in\mathbb{N}$ and  $t\in\mathbb{Z}$.  

%
%

Under these assumptions, denote by $\mathcal{H}^{\bf X}$ the Hilbert space spanned by $\bf X$, equipped with the~L$_2$ covariance scalar product, that is, the set of all L$_2$-convergent linear combinations of~$X_{it}$'s and  limits of L$_2$-convergent sequences thereof. Similarly, denote by~$\mathcal{H}^{\bf X}_t$,  $\mathcal{H}^{{\bf X} ^{(n)}}\!$, and~$\mathcal{H}^{{\bf X} ^{(n)}}_t$  the   subspaces of $\mathcal{H}^{\bf X}$ spanned , respectively, by~$\{{ X}_{is}\vert\ i\in\mathbb{N},\  s\leq t\}$, by~$\{{ X}_{is}\vert\ 1\leq i\leq n, \   t\in\mathbb{Z}\}$, and by~$\{{ X}_{is}\vert \ 1\leq  i\leq n, \ s\leq t\}$. 
 Let~$\eta_0:=\sum_{i=1}^\infty \sum_{s=-\infty}^\infty a_{is}X_{is}\in\mathcal{H}^{\bf X}$. Then, 
 $$
 \eta_t:=\sum_{i=1}^\infty \sum_{s=-\infty}^\infty a_{i,s+t}X_{i,s+t}\in\mathcal{H}^{\bf X}\quad \text{  for all~$t\in\mathbb{Z}$},$$
 and we say that the process ${\pmb\eta}:=\{\eta_t\vert\, t\in\mathbb{Z}\}$ belongs to  $\mathcal{H}^{\bf X}$. 
 
 The main idea in {Hallin et al.~(2000)  and Forni and Lippi~(2001) consists in decomposing $\mathcal{H}^{\bf X}$ into two mutually orthogonal subspaces $\mathcal{H}^{\bf X}_{\text{\rm com}}$ and $\mathcal{H}^{\bf X}_{\text{\rm idio}}\coloneqq \left(\mathcal{H}^{\bf X}_{\text{\rm com}}\right)^\perp$ where  $\mathcal{H}^{\bf X}_{\text{\rm com}}$ denotes the subspace spanned by the limits of all sequences of\linebreak standardized  linear combinations $\sum_{i=1}^n\sum _{k\in{\mathbb Z}} a_{ik} X_{i, t-k}$ of the past, present, and future values of~$X_{it}$'s with squared coefficients summing up to one exhibiting  exploding  variances as $n\to\infty$. 
 
 More precisely, call  \textit{common} any random variable $\zeta$    
  in ${\mathcal H}^{\bf X}$ with variance $0<\sigma^2_\zeta$ such that $\zeta/\sigma_\zeta$ is the limit  in quadratic mean of a sequence $w_{\bf X}
  \n/(\text{Var}(w
  _{\bf X}
  \n))^{1/2}$   of standardized elements of ${\mathcal H}^{\bf X}$  of the form   $$w_{\bf X}
  \n:=\sum_{i=1}^n\sum_{k=-\infty}^\infty a_{ik}^{(n)}X_{i,t-k}\quad\text{  with }\quad \sum_{i=1}^n\sum_{k=-\infty}^\infty (a_{ik}^{(n)})^2 =1$$  such that $\lim_{n\to\infty}\text{Var}(w_{\bf X}
  \n)=\infty$.


 Define the Hilbert space  $\mathcal{H}_{\text{\rm com}}^{\bf X}$ spanned by the collection of all common variables in ${\mathcal H}^{\bf X}$  and  its orthogonal complement (with respect to ${\mathcal H}^{\bf X}$) $\mathcal{H}_{\text{\rm idio}}^{\bf X}:=\big(\mathcal{H}_{\text{\rm com}}^{\bf X}\big)^\bot$ as~$\bf X$'s \textit{common} and \textit{idiosyncratic spaces}, respectively. A process $\bf X$ is called \textit{purely common} if  $\mathcal{H}^{\bf X}=\mathcal{H}_{\text{\rm com}}^{\bf X}$ (hence, $\mathcal{H}_{\text{\rm idio}}^{\bf X}=\{0\}$), \textit{purely idiosyncratic} if $\mathcal{H}^{\bf X}=\mathcal{H}_{\text{\rm idio}}^{\bf X}$  (hence,~$\mathcal{H}_{\text{\rm com}}^{\bf X}=\{0\}$).

 Projecting each $X_{it}$ onto  $\mathcal{H}^{\bf X}_{\text{\rm com}}$ and its orthogonal complement $\mathcal{H}^{\bf X}_{\text{\rm idio}}$ yields the factor model decomposition
 \begin{equation}\label{GDFMdef}X_{it} = \chi_{it} + \xi_{it},\quad i\in\mathbb N,\ \ t\in\mathbb{Z}
 \end{equation}
of $X_{it}$ into a {\it common component} $\chi_{it}$ and an {\it idiosyncratic} component $\xi_{it}$, respectively: call \eqref{GDFMdef} the {\it General Dynamic Factor (GDFM) representation} of $\bf X$. 

Contrary to the factor model decompositions \eqref{1}, \eqref{2}, and \eqref{3} previously considered, the GDFM decomposition \eqref{GDFMdef} is {\it endogenous}; it  always exists, and does not impose any restriction (beyond second-order statio\-narity) on the data-generating process of ${\bf X}$. In that sense,  \eqref{GDFMdef} is not a statistical {\it model}, but a representation result. Whether  it constitutes the description of a data-generating process or not is relevant to the analysis. This representation result nature of \eqref{GDFMdef} was first emphasized in Forni and Lippi~(2001) where, however, a frequency domain approach is adopted.

So far, indeed, no assumption has been imposed on the second-order stationary process  $\bf X$. Adding the requirement that, for any $n\in\mathbb N$, ${\bf X}\n$ admits a 
 {\it spectral density matrix} $\theta\mapsto {\boldsymbol\Sigma}\n({\theta})$, $\theta\in (-\pi,\pi]$ with eigenvalues $\lambda\n_{X;1}(\theta)\ge \lambda\n_{X;2}(\theta)\ge \ldots\ge  \lambda\n_{X;n}(\theta)$ such that 
 \[ 
\lim_{n\to\infty}\lambda\n_{X;q}(\theta)=\infty\quad \text{ and }\quad \lim_{n\to\infty}\lambda\n_{X;q+1}(\theta)  <\infty, \quad \theta\text{-a.e. in }(-\pi, \pi],
\]
{for some finite  $q\in{\mathbb N}$ independent of $n$,} 
it can be shown (see Hallin and Lippi~(2013)) that all $\{\chi_{it} \vert t\in\mathbb{Z}\}$'s are driven by a~$q$-dimensional orthonormal white noise process~$\{{\bf u}_t = (u_{1t} , \ldots , u_{qt})\pr \vert t\in\mathbb{Z}\}$ of  {\it common shocks}.  
The GDFM decomposition~\eqref{GDFMdef}, in that case, takes the form 
 \begin{equation}\label{GDFMshocks}X_{it} = \chi_{it} + \xi_{it}\eqqcolon \sum_{i=1}^q {\bf B}_i(L)
{\bf u}_t  + \xi_{it}
 \quad i\in\mathbb N,\ \ t\in\mathbb{Z}
 \end{equation}
 for  some  collection $ {\bf B}_i(L)\coloneqq \left(
B_{i1}(L),\ldots , B_{iq}(L) \right)
$ of  one-sided linear $1\times q$  square-summable  filters  $B_{ij}(L) \coloneqq  \sum_{k=0}^\infty B_{ijk}L^k$, $i\in\mathbb{N},\ j=1,\ldots , q$.   We refer to Section~3.3 of Hallin and Lippi~(2013) for the relation between the spectral density matrices~$\Sigmab\n(\theta)$ and the filters $ {\bf B}_i(L)$.}

It follows from the above results  that the GDFM (loading filters and factors) is asymptotically identified as $n\to\infty$.  {Forni et al. (2000) show that the common and idiosyncratic  components $\chi_{it}$ and $\xi_{it}$ can be consistently estimated, as $n,T\to\infty$, via dynamic (spectral) PCA, a technique introduced by Brillinger (2001) which, unfortunately, involves two-sided filters, hence performs poorly at the ends of the observation period---making it unsuitable in the context of prediction roblems. 
Forni et al. (2017) show that also the loadings and the factors can be consistently estimated, as $n,T\to\infty$, via a multi-step approach based on an equivalent autoregressive repre\-sentation (not presented here) derived from the results by Anderson and Deistler (2008) and Forni et al. (2015) on singular stochastic processes. See also the recent results by Barigozzi et al. (2023). This latter approach only involves one-sided filters, and  allows for constructing GDFM-based forecasts. Forni et al. (2018) show that such forecasts improve over the Stock and Watson (2002a) ones based on the static factor model.}


\subsection{ Restricted General Dynamic Factor Model: Forni, Hallin, Lippi, and Reichlin (2005) and Forni, Giannone, Lippi, and Reichlin (2009)}

 Forni et al. (2005) consider a restricted version of the GDFM where it is assumed that the infinite singular moving average representation for the common component in \eqref{GDFMshocks} is in fact finite with maximum lag equal to $s$, say. In this case, the GDFM can be written as a model with static loadings and an $r=q(s+1)$-dimensional common factor process ${\bf F}_t=({\bf u}_t^\prime,\ldots, {\bf u}_{t-s}^\prime )^\prime$. 
Note that under this model we must have $q\le r$, a finding often supported by data, see, e.g, the evidence in D'Agostino and Giannone~(2012). The lagged values ${\bf u}_{t-1},\ldots,{\bf u}_{t-s}$ of the ``original factors,'' moreover, should satisfy the pervasiveness conditions leading to condition~(iii) of Chamberlain's {\it approximate static} factor model. 

This \textit{restricted General Dynamic Factor Model} is also often called the \textit{approximate dynamic} factor model where, however, ``dynamic'' refers to the nature of the factors rather than to their loadings, which are static. 
Estimation is typically by dynamic PCA to recover the common component spectral density matrix
plus classical (static) PCA on the recovered common component covariance matrix. This approach is particularly successful in forecasting and the construction of coincident indicators of economic activity, as, e.g, the EuroCoin indicator by Altissimo et al. (2010).

Forni et al.~(2009) further assume that the factors follow a VAR process. It then becomes immediately clear that {such  a VAR has to be singular as soon as $s>0$, i.e., the innovations must have dimension $q<r$. For more details, see the survey by Stock and Watson (2016). Clearly, this approach is almost equivalent to the state-space approach described in Section~4.1, with the only non-trivial difference that now the VAR for the factors is singular.}
Estimation is typically by PCA plus classical estimation of a VAR on the resulting estimators of the latent factors ${\bf F}_t$. Applications are in the macroeconometrics field of impulse response analysis, where the common shocks ${\bf u}_t$ (or an orthogonal transformation thereof) are identified as sources of economic fluctuations, see, e.g., Bernanke et al. (2005).


\section{Identifying the number of factors}\label{SecIdent} 


Irrespective of the choice of a factor model and the  estimation method adopted, identifying the number $r$ of factors in the static loadings case and/ the number $q$ of common shocks in the GDFM is a crucial  preliminary step. A number of methods have been considered. 

The first one,  based on information criteria, was proposed by Bai and Ng (2002) to determine $r$ in the approximate static model, followed  by Hallin and Li\v ska~(2007)  to determine $q$ in the GDFM. The latter work
also proposes  a tuned penalty version of the information criterion method.  Back to the static model, Alessi et al.~(2010) improve on Bai and Ng by combining their criterion with the same tuning idea. 
 {In a restricted GDFM setting, Amengual and Watson~(2007) are adapting the Bai and Ng procedure to  determining $q$, while Bai and Ng (2007) propose a way to jointly estimate $r$ and $q$.}

Another strand of methods is based on the empirical distribution of the eigenvalues of the covariance matrix of the observations. 
Ahn and Horestein~(2007) propose an ``eigenvalue ratio''  and a ``growth ratio'' criterion. The eigenvalue ratio criterion consists in selecting as the number $r$ of factors the number $\hat r$ that maximizes the ratio between the $k$th and the $(k+1)$th eigenvalues arranged in decreasing order of magnitude---a  variant of the  classical (and often decried) {\it scree test} method searching for a ``clear break'' in the spectrum. The growth ratio criterion proceeds similarly, now with the growth rates of the idiosyncratic variances associated with the fitting of $k$ factors. More recently, Avarucci et al.~(2022)   developed dynamic counterparts, based on the eigenvalues of the spectral density matrix  of the observations, of the eigenvalue ratio and growth ratio estimators of Ahn and Horenstein~(2013). Finally, Onatski~(2010) and Trapani (2018) consider the behavior of the difference between the  the $k$th and the $(k+1)$th eigenvalues, showing that this difference, for $k=r$ diverges to infinity while it converges to zero for any $k$ larger than $r$. Onatksi (2009) considers instead the asymptotic behavior of the eigenvalues of the spectral density matrix of the observations in order to determine $q$.

Popular factor number estimators, however, often suffer from the lack of significant eigengap
in empirical eigenvalues and tend to over-estimate $r$ due, for example, to the
existence of non-pervasive factors affecting only a subset of the series, or the presence of moderate cross-sectional correlations
in the idiosyncratic components. Barigozzi and Cho (2020)
show how such overestimation can compromise the consistency of the principal component estimator in the approximate static factor model. They also propose a remedy involving a modified principal component estimator based on a rescaling of the sample eigenvector entries; this modified estimator is shown to be robust against the overestimation of $r$.


\section{`` ... as the stars of the heaven and as the sand which is upon the seashore''\footnote{Genesis 22:17, King James version.}}

The foundational contributions described in Section~\ref{sec3} have triggered a veritable  explosion of further papers on the subject, some of them refining or extending, some others developing related problems, along with countless applications in a variety of fields---much beyond econometrics and finance.  Dynamic factor models, indeed, have emerged as a successful and widely used tool for analyzing the information contained in observed  high-dimensional  time-series (large panels  of time series data), thereby obtaining now-casts and short-term forecasts of economic activity, financial volatility,  inflation, etc. Factor models are used also in environmental and climatic sciences (Marotta and Mumtaz 2023), in health and biomedical studies (Peracchi and Rossetti~2022), and even, back to the origins of the method, in psychology (Molenaar and Ram~2009). 

A Google search on ``Dynamic Factor Model''  brings no less than 435 million entries---{\it as many ``as the stars of the heaven and as the sand which is upon the seashore!''} Below is a short personal, obviously highly  incomplete,  and unavoidably biased list of some of the uncountable offsprings of Stock and Watson (2002), Bai~(2003), and Forni et al. (2000); we regroup them by subject.


\begin{enumerate}
\item[]{\bf Factor models and volatilities}. Factor models as described in the previous sections only are dealing with unconditional covariances. In many applications---certainly so in finance---conditional covariances and volatilities are at least as important. Therefore, many papers have been devoted to volatilities and their forecasts in factor models. The basic and natural  idea is a decomposition of volatilities into common and idiosyncratic. Some authors  (Ng et al.~1992; Harvey et al.~1992;  Connor et al.~2006;  Sentana et al.~2008;  Fan et al.~2015) are considering the volatility of the common components as the common volatility, neglecting as idiosyncratic the volatility of the idiosyncratic components. A different point of view is adopted in Barigozzi and Hallin (2016, 2017a, 2020) 
 and    Truc\`\i os et al.~(2022), where it is argued that the volatility of an  idiosyncratic  component, for instance, may well be exposed to common volatility shocks, and the volatility of a common component be affected by an idiosyncratic volatility shock, so that none of them should be neglected in the forecast exercise. This is used, e.g., in Hallin and Truc\`\i os~(2022), to produce forecasts of Value-at-Risk and Expected Shortfall of large portfolios.\\ 
 \item[] {\bf Factor models and robustness}. Factor model methods  remain  second-order ones: being based on second-
 order dependence structures,they are bound to be sensitive to the possible presence of
 outliers. 
  Many outlier detection procedures  are available in the time series 
 literature,   most of them restricted to univariate series. Relatively little attention has been given to robustness issues in the context of high-dimensional
 time series and factor models, though, with  a handful of references   such as
 Kristensen (2014) and Baragona and Battaglia~(2007), who show that both the traditional
 (static) PCA methods and the more general dynamic PCA methods yield biased
 estimates in the presence of outliers; more recent contributions include Fan et al.~(2018, 2019), Truc\`\i os et al.~(2019), Alonso et al. (2020), and He et al.~(2022, 2023).   The robustness of the  Forni et al. (2015, 2017) approach is investigated in Truc\`\i os et al.
 (2021).\\  
\item[]{\bf Factor models with blocks and hierarchical factor models}. A panel often decomposes into subpanels or blocks, according, for instance, to geographic criteria. Such features  are treated via a  (finite-$n$, large $T$)  hierarchical static factor model  by  Kose et al.~(2003)  and (large $n$ and $T$) Moench et al.~(2013), in full generality and a (large $n$ and $T$)  GDFM by  Hallin and Li\v ska (2011), Hallin et al. (2011), and  Barigozzi et al. (2018b).\\ 
\item[]{\bf Locally stationary factor models}. In practice, time series observed over a long period of time $T$ seldom are stationary; the evolution over  time of the data-generating process, however, often can be considered to be smooth. This, in the univariate case, has motivated the development, initiated by Dahlhaus~(1997), of the so-called {\it locally statio\-nary} approach  where it is assumed that the second-order structure is evolving slowly over time. 
 Barigozzi et al.~(2021a) consider  the estimation of a time-varying version of the GDFM in which the factors are loaded via time-varying filters.   A slightly different  time-varying version of the GDFM,  inspired by Forni et al.~(2000),  had been previously studied by Eichler et al.~(2011) which, however,   is entirely based on Brillinger's dynamic principal component analysis, hence suffers from the main drawback of dynamic principal components, which resorts to  two-sided filters to recover the space spanned by the factors. Such two-sided filtering    makes the  Eichler et al.~(2011)  approach unsuitable for  forecasting and  impulse response analysis. This two-sidedness issue, in  Barigozzi et al.~(2021a), is taken care of via the one-sided approach developed in Forni et al.~(2015, 2017). An approximate static version with time-varying loading matrices  is considered in Hafner et al.~(2011).\\ 
 \item[]{\bf Integrated factors}. Bai and Ng (2004) and Barigozzi et al. (2021b) extend the static factor model to allow for the presence of unit roots,  both in the factors and in the idiosyncratic processes. While Bai and Ng (2004) develop a testing procedure for the presence of unit roots, and Barigozzi et al (2020, 2021b) 
assume the joint presence of unit roots jointly and  deterministic trending components and study principal-component-based estimation as well as the estimation of a vector error correction model for the  factors. This allows for computing impulse response functions in presence of permanent and transitory shocks. It is important to notice that a realistic factor model must allow for the idiosyncratic component to be integrated too---unless we assume the  $n$ observed  series to be cointegrated with cointegration rank $(n-r)$, where $r$ is the number of factors which, in this case, are also common trends (see Bai (2004) and Barigozzi and Trapani (2022)).\\

\item[]{\bf Functional factor models}. 
Gao et al.~(2019) consider the problem of forecasting
 high-dimensional functional time series via  a heuristic two-stage approach combining a truncated-PCA dimension reduction and separate scalar factor-model analyses of the resulting (scalar) panels, conducted via an eigendecomposition of the long-run covariance operator, as opposed to the lag-zero covariance operator considered in static factor models;  the
truncation of the PCA decomposition, moreover, potentially may lead to a dramatic loss of information. In Gao et al.~(2021), the same authors propose a factor-augmented version of their approach. 
 Tang et al.~(2021) propose a functional factor model allowing both factors and
loadings to be functional,  Guo et al.~(2021)   a functional factor model with
functional factors and scalar loadings. In both cases, the model is considered as the description of a very specific data-generating process (no representation result), hence    requires  being checked;    all the component functional time
series, moreover, have to take values in the same Hilbert space, which is somewhat  restrictive. Hallin et al. (2023); Tavakoli et al.~(2023), on the other hand, extend to the functional case the approximate static model of Chamberlain, and establish a representation result; the component time series may take values in different Hilbert spaces, including the real line.\\ 
\item[]{\bf Factors plus networks}. Among other possible dimension reduction techniques, sparse regressions, as lasso or ridge, are the main competitors of factor analysis which favors instead a dense modelling of the data. A crucial question to be asked in empirical analysis  then is whether   a given dataset is sparse or dense. It has been shown empirically that most economic time series datasets have a dense structure, so factor analysis should be preferred as it is likely to deliver better forecasts (De Mol et al.~2008). On the other hand, especially in financial data, once we control for common factors, there is evidence of non-negligible dependencies left in the idiosyncratic components (Barigozzi and Hallin 2017b).
The theoretical properties of a factor plus sparse approach are studied by Fan et al. (2023) in the static loadings framework and by Barigozzi et al. (2023) in the GDFM framework. Idiosyncratic components  there are modelled as  a sparse VAR and the estimated coefficients, which are sparse matrices, have often been given a network interpretation in which edges represent non-zero conditional correlations.\\
\item[]{\bf  Factors and breaks}. The presence of structural breaks or change-points represents another important cause of deviation from the assumption of stationarity. Many papers have investigated this problem from an off-line perspective in the static factor model approach, proposing various tests for the presence of breaks and, possibly, estimators for their location: see 
Breitung and Eickmeier (2011), 
Chen et al. (2014), 
Han and Inoue (2014), 
Corradi and Swanson (2014), 
Yamamoto and Tanaka (2015), 
Cheng et al. (2016), 
Baltagi et al.~(2017),  
Bai et al.~(2017), 
Ma and Su (2018), 
Duan et al (2023).  
The case of the restricted dynamic factor model has been studied by Barigozzi et al.~(2018a) who not only allow  for changes in the loadings, but also in the number of factors, and in the autocorrelation function of the factors, as well as change-points in the idiosyncratic  second-order structure. A similar approach but for the GDFM setting is in Cho et al. (2023). Finally, the issue of on-line, i.e., sequential, testing for change-points in factor models which, despite of its importance for practice, has not received much attention, is considered in Barigozzi and Trapani (2020).
\end{enumerate}

\section*{Acknowledgments}
Marc Hallin gratefully acknowledges the support of the Czech Science Foundation grant GA\v{C}R22036365.


\begin{thebibliography}{9} 

\bibitem[Ahn and Horenstein~(2013)]{} S.C.\ Ahn and A.R.\ Horenstein~(2013), ``Eigenvalue ratio test for the number of factors,''  {\it Econometrica}, Vol. 81, 1203--1227.

\bibitem[Alessi et al.~(2010)]{Alessetal10}L.\  Alessi, M.\ Barigozzi, and M.\ Capasso~(2010), ``Improved penalization for determining
 the number of factors in approximate factor models,''   {\it Statistics \& Probability Letters}, Vol.\ 
 80, pp.\ 1806--1813.
 
 \bibitem[Alonso et al.~(2020)]{}, A. M.~Alonso, P.~Galeano,  and D.~Pe{\~n}a~(2020), ``A robust procedure to build dynamic factor models with cluster structure,'' {\it Journal of Econometrics}, Vol.~216, pp.~35--52.
 
 
\bibitem[Altissimo et al. (2010)]{ACFLV10} F. \ Altissimo, R.\ Cristadoro, M.\ Forni, M. \ Lippi, and G. Veronese~(2010), ``New Eurocoin: tracking economic growth in real time,''  {\it The Review of Economics and Statistics}, Vol.\ 92, pp.\ 1024--1034.

 \bibitem[Amemiya et al.~(1987)]{AFP87} Y. \ Amemiya, W.A.\ Fuller, and S.G. \ Pantula~(1987), ``The asymptotic distributions
of some estimators for a factor analysis model,''  {\it Journal of Multivariate Analysis}, Vol.\ 22, pp.\ 51--64.
 
  \bibitem[Amengual and Watson~(2007)]{AW07} D.\ Amengual and M.W. Watson~(2007), ``Consistent estimation of the number of dynamic factors in a large $N$ and $T$ panel,''  {\it Journal of Business \& Economic Statistics}, Vol.\ 25, pp.\ 91--96.

 \bibitem[Anderson and Rubin (1956)]{AR56} T.W.\ Anderson and H.\ Rubin~(1956), ``Statistical inference in factor analysis,''  in: J.\ Neyman, Ed., {\it Proceedings of the Third Berkeley Symposium on Mathematical Statistics and Probability}, Vol.~1, pp.\ 111--150, University of California Press.

\bibitem[Anderson and Deistler (2008)]{Anderson and Deistler (2008)}B.D.O.\ Anderson and M.\ Deistler (2008), ``Properties of zero-free transfer function matrices,''  {\it SICE Journal of Control, Measurement and System Integration}, Vol.~1, pp.\  284--92. 

%
%





  \bibitem[Avarucci et al.~(2022)]{ACFZ22}
  M.\ Avarucci, M.\ Cavicchioli, M.\ Forni, and P.\ Zaffaroni~(2022),  ``The main business cycle shock(s): frequency-band estimation of the number of dynamic factors,''  CEPR DP17281.

 \bibitem[Bai~(2003)]{Bai03}J.\ Bai~(2003), ``Inferential theory for factor models of large dimensions,''   {\it Econometrica}, Vol.\  71, pp.\ 135--171.
 
 \bibitem[Bai (2004)]{} J.\ Bai~(2004), ``Estimating cross-section common stochastic trends in nonstationary panel data,'' {\it Journal of Econometrics}, Vol.\ 122, pp.\ 137--183.
 
 \bibitem[Bai et al. (2020)]{} J.\ Bai, X.\ Han, and Y. Shi~(2020), ``Estimation and inference of change points in high-dimensional factor models,''  {\it Journal of Econometrics}, Vol.\ 219, pp.\ 66--100.

 
  \bibitem[Bai and Ng~(2002)]{BaiNg02}J.\ Bai and S.\ Ng~(2002), ``Determining the number of factors in approximate factor models,''   {\it Econometrica}, Vol.\  70, pp.\ 191--221.

  \bibitem[Bai and Ng~(2004)]{BaiNg04}J.\ Bai and S.\ Ng~(2004), ``A PANIC attack on unit roots and cointegration,''   {\it Econometrica}, Vol.\  72, pp.\ 1127--1177.
  
    \bibitem[Bai and Ng~(2006)]{BaiNg06}J.\ Bai and S.\ Ng~(2006), ``Confidence intervals for diffusion index forecasts and inference for factor-augmented regressions,''   {\it Econometrica}, Vol.\  74, pp.\ 1133--1150.
  
    \bibitem[Bai and Ng~(2007)]{BaiNg07}J.\ Bai and S.\ Ng~(2007), ``Determining the number of primitive shocks in factor models,''  {\it Journal of Business \& Economic Statistics}, Vol.\ 25, pp.\ 52--60.
  
  \bibitem[Bai and Li~(2012)]{BaiLi12}J.\ Bai and K.\ Li~(2012), ``Statistical analysis of factor models of high dimension,''  {\it The Annals of Statistics}, Vol.\ 40, pp.\ 436--465.

  \bibitem[Bai and Li~(2016)]{BaiLi16}J.\ Bai and K.\ Li~(2016), ``Maximum likelihood estimation and inference for approximate factor models of high dimension,''   {\it The Review of Economics and Statistics}, Vol.~98, pp.\ 298--309.
  
  \bibitem[Baltagi et al. (2017)]{}B.H.\ Baltagi, C.\ Kao, and F.\ Wang (2017) ``Identification and estimation of a large factor model with structural instability,'' {\it Journal of Econometrics}, Vol.\ 197, pp. 87--100.

  
  \bibitem[Baragona and Battaglia~(2007)]{}  R.~Baragona and  F.~Battaglia (2007), ``Outliers in dynamic factor models,'' 
{\it Electronic Journal of Statistics}, Vol.~1, pp.~392--432.

 
 \bibitem[Barigozzi and Cho (2020)]{BC20} M.\ Barigozzi and H.\ Cho~(2020), ``Consistent estimation of high-dimensional factor models when the factor number is over-estimated'', {\it Electronic Journal of Statistics}, Vol.\ 14, pp.\  2892--2921.

 
 \bibitem[Barigozzi et al. (2018a)]{}M.\ Barigozzi, H.\ Cho, and P.\ Fryzlewicz~(2018), ``Simultaneous multiple change-point and factor analysis for high-dimensional time series'', {\it Journal of Econometrics}, Vol.\ 206, pp.\ 187--225.

  \bibitem[Barigozzi et al. (2023)]{BCO23} M.\ Barigozzi, H.\ Cho, and D.\ Owens~(2023), ``FNETS: factor-adjusted network estimation and forecasting for high-dimensional time series,'' {\it Journal of Business \& Economic Statistics}, to appear.


 \bibitem[Barigozzi and Hallin~(2016)]{BH16}M.\ Barigozzi and M.\ Hallin~(2016), ``Generalized dynamic factor models and volatili\-ties:\  recovering the market volatility shocks,''  {\it The Econometrics Journal}, Vol.\ 201, pp.\ 307--321.
 
  \bibitem[Barigozzi and Hallin~(2017a)]{BH17}M.\ Barigozzi and M.\ Hallin~(2017a), ``Generalized dynamic factor models and volati\-lities: estimation and forecasting,''  {\it Journal of Econometrics}, Vol.\ 201, pp.\ 33--60.
  
  \bibitem[Barigozzi and Hallin (2017b)]{BH17} M.\ Barigozzi and M.\ Hallin~(2017b), ``A network analysis of the volatility of high-dimensional financial series'', {\it Journal of the Royal Statistical Society: Series C (Applied Statistics)}, Vol.\  66, pp.\  581--605.

  \bibitem[Barigozzi and Hallin~(2020)]{BH20}M.\ Barigozzi and M.\ Hallin~(2020), ``General dynamic factor models and volatilities: consistency, rates, and prediction intervals,''  {\it Journal of Econometrics}, Vol.\ 216, pp.\ 4--34.

 \bibitem[Barigozzi et al.~(2023)]{Barigoetal23}M.\ Barigozzi, M.\ Hallin, M.\ Luciani, and P. Zaffaroni~(2023), ``Inferential theory for Generalized Dynamic Factor Models,''   {\it Journal of Econometrics}, 
 to appear. 
  
  \bibitem[Barigozzi et al.~(2018b)]{BHS18b} 
  M.\ Barigozzi, M.\ Hallin, and S.\ Soccorsi~(2018b), ``Identification of global and local shocks in international financial markets via general dynamic factor models”,  {\it Journal of Financial Econometrics}, Vol.\ 17, pp.\ 462--494.

  \bibitem[Barigozzi et al.~(2021a)]{BHSvS21} 
   M.\ Barigozzi, M.\ Hallin, S.\ Soccorsi, and R.\ von Sachs~(2021a), ``Time-varying general dynamic factor models and the measurement of financial connectedness”, {\it Journal of Econometrics}, Vol.\ 222, pp.\ 324--343.
   
   \bibitem[Barigozzi et al (2020)]{}M.\ Barigozzi, M.\ Lippi, and M.\ Luciani (2020), ``Cointegration and error correction mechanisms for singular stochastic vectors,''  {\it Econometrics}, Vol.\ 8, pp.\ 1--23.

  
    \bibitem[Barigozzi et al.~(2021b)]{BLL21}
    M.\ Barigozzi, M.\ Lippi, and M.\ Luciani~(2021),
    ``Large-dimensional dynamic factor models: estimation of impulse-response functions with $I(1)$ cointegrated factors,''  {\it Journal of Econometrics}, Vol.\ 221, pp.\ 455--482.
    
        \bibitem[Barigozzi and Luciani~(2021)]{BL21}M.\ Barigozzi and M.\ Luciani~(2021), ``Measuring the output gap using large datasets,''  {\it The Review of Economics and Statistics}, to appear.
        
        \bibitem[Barigozzi and Trapani (2020)]{}M.\ Barigozzi and L.\ Trapani~(2020), ``Sequential testing for structural stability in approximate factor models,'' {\it Stochastic Processes and their Applications}, Vol.\ 130, pp.\ 5149--5187.

        \bibitem[Barigozzi and Trapani (2022)]{}M.\ Barigozzi and L.\ Trapani~(2020), ``Testing for common trends in non-stationary large datasets,''  {\it Journal of Business \& Economic Statistics}, Vol.\  40, pp.\ 1107--1122.

         
 \bibitem[Bartlett (1937)]{bart37} M.S.\ Bartlett~(1937), ``The statistical conception of mental factors,''  {\it British Journal of Psychology}, Vol.\ 28, pp.\ 97--104.
 
 \bibitem[Bartlett (1938)]{bart38} M.S.\ Bartlett~(1938), ``Methods of estimating mental factors,''  {\it Nature}, Vol.\ 141, pp.\ 609--610.
 
   \bibitem[Bernanke et al. (2005)]{BBE05}B.S.\ Bernanke, J.\ Boivin, and P.\ Eliasz~(2005), ``Measuring the effects of monetary policy a FAVAR approach,''  {\it The Quarterly Journal of Economics}, Vol.\ 102, pp.\ 387--422.
   
   \bibitem[Breitung and Eickmeier (2011)]{}J.\ Breitung and S.\ Eickmeier~(2011), ``Testing for structural breaks in dynamic factor models,'' {\it Journal of Econometrics}, Vol.\ 163, pp.\ 71--84.

 
      \bibitem[Brillinger (2001)]{Brill01} D.R.\ Brillinger~(2001), {\it Time series: data analysis and theory}, Society for Industrial and Applied Mathematics.

 \bibitem[Chamberlain (1983)]{Chamberlain83}  G.\ Chamberlain (1983),  ``Funds, factors and diversification in arbitrage pricing models,''   {\it Econometrica}, Vol.\  51, pp.\ 1305--1323.

 \bibitem[Chamberlain and Rothschild (1983)]{ChambRoth83} G.\ Chamberlain and M.\  Rothschild (1983), `` Arbitrage, factor structure, and mean-variance analysis on large asset markets,''    {\it Econometrica}, Vol.\  51, pp.\  1281--1323.
 
 \bibitem[Chen et al. (2014)]{}L.\ Chen, J.J.\ Dolado, and J.\ Gonzalo~(2014), ``Detecting big structural breaks in large factor models,'' {\it Journal of Econometrics}, Vol.\ 180, pp.\ 30--48.
 
 \bibitem[Cheng et al. (2016)]{}X.\ Cheng, Z.\ Liao, and F.\ Schorfheide~(2016), ``Shrinkage estimation of high-dimensional factor models with structural instabilities,'' {\it Review of Economic Studies}, Vol.\ 83, pp.\ 1511--1543.

\bibitem[Cho et al. (2023)]{}H.\ Cho, H.\ Maeng, I.A.\ Eckley, and P. Fearnhead (2023) ``High-dimensional time series segmentation via factor-adjusted vector autoregressive modelling,''  {\it Journal of the American Statistical Association}, to appear.

 
 \bibitem[Connor and Korajczyck (1986)]{CK86} G.\ Connor and R.A.\ Korajczyck~(1986), ``Performance measurement with the arbitrage pricing theory: a new framework for analysis,''  {\it Journal of Financial Economics}, Vol.\ 15, pp.\ 373--394.
 
  \bibitem[Connor et al.~(2006)]{} G.~Connor, R.A.~Korajczyk,  and O.~Linton~(2006), ``The common and specific components of dynamic volatility,''  {\it Journal
of Econometrics}, Vol.~132, pp.~231--255.

\bibitem[Corradi and Swanson (2014)]{}V.\ Corradi and N. Swanson (2014), ``Testing for structural stability of factor augmented forecasting models,''  {\it Journal of Econometrics}, Vol.\ 182, pp.\ 100--118.

 
 
  \bibitem[D'Agostino and Giannone (2012)]{DG12} A.\ D'Agostino and D. Giannone~(2012), ``Comparing alternative predictors based on large-panel factor models,''  {\it Oxford Bulletin of Economics and Statistics}, Vol.\ 74, pp. 306--326.
  

  \bibitem[Dahlhaus (1997)]{}R.~Dahlhaus~(1997), ``Fitting time series models to nonstationary processes,''  {\it The Annals of Statistics}, Vol.~25, pp.~1--37.
  

   \bibitem[De Mol et al. (2008)]{DGR08} C.\ De Mol, D.\ Giannone, and L.\ Reichlin~(2008), ``Forecasting using a large number of predictors: is Bayesian shrinkage a valid alternative to principal components?'' {\it Journal of Econometrics}, Vol.\ 146, pp.\ 318--328. 

  \bibitem[Doz et al. (2011)]{DGR11} C.\ Doz, D.\ Giannone, and L. \ Recihlin~(2011), ``A two-step estimator for large approximate dynamic factor models based on Kalman filtering,''  {\it Journal of Econometrics}, Vol.~164, pp.\ 188--205.

  \bibitem[Doz et al. (2012)]{DGR12} C.\ Doz, D.\ Giannone, and L. \ Recihlin~(2012), ``A quasi-maximum likelihood approach for large, approximate dynamic factor models,''  {\it The Review of Economics and Statistics}, Vol.\ 94, 1014--1024.
  
   \bibitem[Duan et al. (2023)]{}J.\ Duan, J.\ Bai, and X.\ Han~(2023), ``Quasi-maximum likelihood estimation of break point in high-dimensional factor models,''  {\it Journal of Econometrics}, Vol.\ 233, pp.\ 209--236.

   \bibitem[Eichler et al.~(2011)]{} M.~Eichler, G.~Motta, and R.~von Sachs~(2011) ``Fitting dynamic factor models to non-stationary time series,''  {\it Journal of Econometrics}, Vol.~163, pp.~51--70.
   
     \bibitem[Fan et al.~(2015)]{} J.~Fan, Y.~Liao, and X.~Shi~(2015), ``Risks of large portfolios,'' {\it Journal of Econometrics}, Vol.~186, pp.~367--387.

   
   \bibitem[ Fan et al.~(2018)]{}J. Fan, H.~Liu, and W.~Wang~(2018), ``Large covariance estimation through elliptical factor models,''     {\it The Annals of Statistics}, Vol.~46, pp.~1383--1414.
   
   \bibitem[Fan et al. (2023)]{FMM23} J.\ Fan, R.\ Masini, and M.C.\ Medeiros~(2023), ``Bridging factor and sparse models,''  {\it The Annals of Statistics}, to appear.
   
 

   
      \bibitem[ Fan et al.~(2018)]{}J. Fan, W.~Wang, and Y.~Zhong~(2019), ``Robust covariance estimation for approximate factor models,'' {\it Journal of Econometrics}, Vol.~208, pp.~5--22. 
      
      \bibitem[Fiorentini et al. (2018)]{FGS18}G.\ Fiorentini, A.\ Galesi, and E.\ Sentana~(2018), ``A spectral EM algorithm for dynamic factor models,''  {\it Journal of Econometrics}, Vol.\ 205, pp.\ 249--279.
  
 
 
   \bibitem[Forni et al. (2009)]{FGLR09} M.\ Forni, D.\ Giannone, M.\ Lippi, and L.\ Reichlin~(2009), ``Opening the black box: structural factor models with large cross sections,''  {\it Econometric Theory}, Vol.\ 25, pp.\ 1319--1347.
 
  \bibitem[Forni et al.~(2000)]{Fornietal00}M.\ Forni, M.\ Hallin, M.\ Lippi, and L. Reichlin~(2000), ``The generalized dynamic factor model: identification and estimation,''  {\it The Review of Economics and Statistics}, Vol.\ 82, pp.\ 540--554.
  
  \bibitem[Forni et al.~(2015)]{Fornietal15}M.\ Forni, M.\ Hallin, M.\ Lippi, and P.\ Zaffaroni~(2015), ``Dynamic factor models with infinite-dimensional factor spaces: one-sided representations,''  {\it Journal of Econometrics}, Vol.\ 185, pp.\ 359--371.
   
  \bibitem[Forni et al.~(2017)]{Fornietal17}M.\ Forni, M.\ Hallin, M.\ Lippi, and P.\ Zaffaroni~(2017), ``Dynamic factor models with infinite dimensional factor space: asymptotic analysis'' {\it Journal of Econometrics}, Vol.\ 199, pp.\ 74--92.
  
    \bibitem[Forni et al.~(2018)]{Fornietal18} M.\ Forni, A.\ Giovannelli, M.\ Lippi, and S.\ Soccorsi~(2018), ``Dynamic factor model with infinite‐dimensional factor space: forecasting,''  {\it Journal of Applied Econometrics}, Vol.\ 33, pp.\ 625--642.
        
     \bibitem[Forni and Lippi~(2001)]{FLippi01}M. Forni and M. Lippi~(2001), ``The generalized dynamic factor model: representation theory,''   {\it Econometric Theory}, Vol.\ 17, pp.\ 1113--1141.
     
\bibitem[Gao et al.~(2019)]{} Y. Gao, H.L.~Shang,  and Y.~Yang~(2019), ``High-dimensional functional time series forecasting:
an application to age-specific mortality rates,'' {\it Journal of Multivariate Analysis},  Vol.~170, pp.~ 232--243.


\bibitem[Gao et al.~(2019)]{} Y. Gao, H.L.~Shang,  and Y.~Yang~(2021), ``Factor-augmented smoothing model for functional
data,''  arXiv:2102.02580.
     
 
 \bibitem[Geweke (1977)]{Geweke77}  J.\ Geweke (1977),  ``The dynamic factor analysis of economic time series,''   in: D.J.\ Aigner and A.S.\ Goldberger, Eds., {\it Latent variables in socio-economic models,} pp.\ 365--383, Amsterdam: North Holland.

\bibitem[Geweke and Singleton (1981)]{GSingle81}  J.\ Geweke and K.J.\ Singleton (1981), ``Latent variable models for time series: a frequency domain approach with an application to the permanent income hypothesis,''  {\it Journal of Econometrics} Vol.\ 17, pp.\ 287--304.

\bibitem[Giannone et al. (2008)]{GRS08}D.\ Giannone, L.\ Reichlin, and D.\ Small~(2008), ``Nowcasting: the real-time informational content of macroeconomic data,''  {\it Journal of Monetary Economics}, Vol.\ 55, pp.\ 665--676.

\bibitem[Guo et al.~(2021)] {Guoetal21}
S.\ Guo, X.\ Qiao, and Q. Wang (2021), 
``Factor modelling for high-dimensional functional time series,''    arXiv:2112.13651.

\bibitem[Hafner et al.~(2011)]{}C. Hafner, G.~Motta, and R.~von~Sachs~(2011), ``Locally stationary factor models: Identification and nonparametric estimation,'' {\it Econometric Theory}, Vol.~27, pp.~1279--1319.

\bibitem[Hallin~(2023)]{Hallin23} M.\ Hallin (2023), ``Manfred Deistler  
and the General-Dynamic-Factor-Model approach 
to the analysis of high-dimensional time series,''   {\it Econometrics}, Vol.\ 10, pp. 1--9.  
 
 \bibitem[Hallin and Lippi~(2013)]{HallinLippi13} M.\ Hallin  and M.\ Lippi~(2013), ``Factor models in high-dimensional time series: a time-domain approach,''   {\it Stochastic Processes and their Applications}, Vol.\ 123, pp.\ 2678--2695.
 
  \bibitem[Hallin and Li\v ska~(2007)]{HallinLiska07} M.\ Hallin  and and R.~Li\v ska~(2007), ``The generalized dynamic factor model:\ determining the number of factors,''   {\it 
  Journal of the American Statistical Association}, Vol.\ 102, pp.\ 603--617.

  \bibitem[Hallin and Li\v ska~(2011)]{HallinLiska11} M.\ Hallin  and and R.~Li\v ska~(2011), ``Dynamic factors in the presence of blocks,''  {\it Journal of Econometrics}, Vol.\ 163, pp.\ 29--41. 

     \bibitem[Hallin et al. (2011)]{HMPV11} M.\ Hallin, C.\ Mathias, H.\ Pirotte, and D. Veredas~(2011), ``Market liquidity as dynamic factors,''  {\it Journal of Econometrics}, Vol.\  163, pp.\ 42--50.
     
     \bibitem[Hallin et al. (2023)]{HNTava23} M.\ Hallin, G.\ Nisol,  and S.\ Tavakoli~(2023), ``Factor models for high-dimensional
functional time series
I: Representation results,''  {\it Journal of Time Series Analysis}, Vol.\ 44, pp.\  578--600. 

    \bibitem[Hallin and Truc\'{\i}os~(2022)]{Htruc22}M. Hallin and C.~Truc\'{\i}os~(2022), ``Forecasting Value-at-Risk and Expected Shortfall in large portfolios:\  a General Dynamic Factor Model approach,''    {\it Econometrics and Statistics}, Vol.\ 27, pp.\  1-15.
    
    \bibitem[Han et al. (2014)]{} X.\ Han and A.\ Inoue~(2014), ``Tests for parameter instability in dynamic factor models,''  {\it Econometric Theory}, Vol.\ 31, pp.\ 1--36.

    \bibitem[Harvey et al.~(1992)]{} A.~Harvey,  E.~Ruiz, and E.~Sentana~(1992), ``Unobserved component time series models with ARCH disturbances,''  {\it Journal
of Econometrics}, Vol.~52, pp.~129--157.

    \bibitem[He et al.~(2022)]{} Y.~He, X.~Kong, L.~Yu, and X.~Zhang~(2022), ``Large-dimensional factor analysis without moment constraints,''  {\it Journal of Business \& Economic Statistics}, Vol.~40, pp.~302--312.
    
        \bibitem[He et al.~(2023)]{} Y.~He, L.~Li,  D.~Liu, and W.X.~Zhou~(2023), ``Huber principal component ana\-lysis for large-dimensional factor models,''   arXiv:2303.02817.
 

    

\bibitem[Hotelling (1933a)]{hotel33a} H.\ Hotelling~(1933a), ``Analysis of a complex of statistical variables into principal components, Part 1,''  {\it Journal of Educational Psychology}, Vol.\ 24, pp.\ 417--441.

\bibitem[Hotelling (1933b)]{hotel33b} H.\ Hotelling~(1933b), ``Analysis of a complex of statistical variables into principal components, Part 2,''  {\it Journal of Educational Psychology}, Vol.\ 25, pp.\ 498--520.

\bibitem[J\" oreskog (1969)]{Jor69}K.G.\  J\" oreskog~(1969), ``A general approach to confirmatory maximum likelihood factor analysis,''  {\it Psychometrika}, Vol.\ 34, pp.\ 183--202.


\bibitem[J\" oreskog (2007)]{Jor07}K.G.\  J\" oreskog~(2007), ``Factor analysis and its extensions,''  in R.\ Cudeck and R.C.\ MacCallum, Eds,  {\it Factor Analysis at 100: Historical Developments and Future Directions}, pp.\ 47-77, Lawrence Erlbaum Associates, Mahwah, N.J.

 \bibitem[Johnstone (2001)]{Johnstone (2001)}I.M.\ Johnstone (2001),   ``On the distribution of the largest eigenvalue in principal components analysis,''   {\it The Annals of Statistics} {Vol.\ 29}, pp.\ 295--327.
 
 \bibitem[Kose et al.~(2003)], M.A.\ Kose, ,C.\ Otrok, and C.H.\ Whiteman (2003.), "International business cycles: world, region, and country-specific factors,''  {\it American Economic Review,} Vol.\ 93,\linebreak pp.~1216-1239. 
 
  \bibitem[Kristensen~(2014)]{} J. T.~Kristensen (2014), ``Factor-based forecasting in the presence of outliers: are factors better selected and estimated by the median than by the mean?''  {\it Studies in Nonlinear Dynamics \& Econometrics},  Vol.~18, pp.~309--338.
 
 
   \bibitem[Lam et al.~(2011)]{LYB11} C.\ Lam, Q.\ Yao, and N.\ Bathia~(2011), ``Estimation of latent factors for high-dimensional time series,''  {\it Biometrika}, Vol.\ 98, pp.\ 901--918.

   \bibitem[Lam and Yao~(2012)]{LamYao12}C.\ Lam and Q.\ Yao~(2012), ``Factor modeling for high-dimensional time series: 
inference for the number of factors,''  {\it The Annals of Statistics}, Vol.\ 40, pp.\ 694--726.
 
  \bibitem[Lawley and Maxwell (1971)]{Lawley71} D.N.\ Lawley  and A.E.\  Maxwell (1971), {\it Factor Analysis as a Statistical Method}, 2nd Edition, Butterworths, London.


\bibitem[Lippi, Deistler, and Anderson~(2023)]{Lippietal23}M.\ Lippi, M.\ Deistler, and B.D.O.\ Anderson~(2023),``High-dimensional dynamic factor models: a selective survey and lines of future research,''  {\it Econometrics and Statistics}, Vol.\ 26, pp.\ 3--16.


\bibitem[Ma and Su (2018)]{}S.\ Ma and L.\ Su~(2018), ``Estimation of large dimensional factor models with an unknown number of breaks,''  {\it Journal of Econometrics}, Vol.\ 207, pp.\ 1--29.

\bibitem[Marotta and Mumtaz~(2023)]{MM23}F.\  Marotta and H.\ Mumtaz~(2023), ``Vulnerability to climate change:
evidence from a dynamic factor model,''  Smith School Working Paper 23-06, University of Oxford. 


 \bibitem[Moench et al.~(2013)]{} E.~Moench, S.\ Ng, and S.\ Potter,~(2013), ``Dynamic hierarchical factor models,'' {\it The Review of Economics and Statistics}, Vol.\ 95, pp.\~1811--1817.


\bibitem[Molenaar and Ram~(2009)]{psycho09}P.CM.\ Molenaar and N. Ram~(2009), ``Advances in dynamic factor analysis of psychological processes,''  in J.\ Valsiner, P.C.M.\  Molenaar, M.C.D.P.\ Lyra, and  N. Chaudhary (Eds.), {\it Dynamic process methodology in the social and developmental sciences}, pp.\  255?268, Springer Science.  

 \bibitem[Onatski (2009)]{Onatski09} A.\ Onatski~(2009), ``Testing hypothesis about the number of factors in large factor models,'' {\it Econometrica}, Vol.\ 77, pp.\ 1447--1479.
 
  \bibitem[Onatski (2010)]{Onatski10} A.\ Onatski~(2010), ``Determining the number of factors from empirical distribution of eigenvalues,''  {\it The Review of Economics and Statistics}, Vol.\ 92, pp.\ 1004--1016.
 
 \bibitem[Onatski et al. (2013)]{Onatski13}A.\ Onatski, M.J.\ Moreira, and M.\ Hallin (2013), ``Asymptotic power of sphericity tests for high-dimensional data,''   {\it The Annals of
Statistics},  Vol.\ 41, pp.~1204--1231.

 \bibitem[Onatski et al. (2014)]{Onatski14}A.\ Onatski, M.J.\ Moreira, and M.\ Hallin (2014), ``Signal detection in high dimension: The multispiked case,''  {\it The Annals of
Statistics},  Vol.\ 42, pp.~225--254.

\bibitem[Pearson~(1901)]{}K.\ Pearson (1901),``On lines and planes of closest fit to systems of points in space,'' {\it Philosophy
Magazine}, Vol. 2, pp.\  559--572.


\bibitem[Pe\~{n}a and Box~(1987)]{PenaBox87}D.\ Pe\~{n}a and G.E.P.\ Box~(1987), ``Identifying a simplifying structure in time series,''  {\it Journal of the American Statistical Association} Vol.\ 82, pp.\ 836--843.

\bibitem[Peracchi and Rossetti~(2022)]{Peracchi22}F.\ Peracchi and C. Rossetti~(2022), ``A nonlinear dynamic factor model of health and medical treatment,''  {\it Health Economics}, Vol.\ 31, pp.\ 1046--1066. 

\bibitem[Poncela et al. (2021)]{PRM21}  P.\ Poncela, E.\ Ruiz, and K.\ Miranda~(2021) ``Factor extraction using Kalman filter and smoothing: this is not just another survey,''  {\it International Journal of Forecasting}, Vol.\ 37, pp.\ 1399--1425.

  \bibitem[Quah and Sargent~(1993)]{QS93} D.\ Quah and T.J.\ Sargent~(1993), ``A dynamic index model for large cross sections,''  in:
  J.H.\ Stock and M.W. Watson, Eds.,
   {\it Business cycles, indicators and forecasting}, pp. 285--306, University of Chicago Press.
%


 \bibitem[Sargent and Sims (1977)]{SSims77}  T.J.\ Sargent and C.A.\ Sims (977),``Business cycle modeling without pretending to have too much a priori economic theory,''  in: C.A.\ Sims,  Ed., {\it New Methods in
Business Cycle Research}, pp.\ 45--109, Federal Reserve Bank of Minneapolis.

 \bibitem[Sentana et al.~(2008)]{}E.~Sentana, G.~Calzolari, and G.~Fiorentini~(2008), ``Indirect estimation of large conditionally heteroskedastic factor models, 
with an application to the Dow 30 stocks,''  {\it Journal of Econometrics}, Vol.~146, pp.~10--25. 
 



\bibitem[Shumway and Stoffer (1982)]{SS82} R.H.\ Shumway and D.S.\ Stoffer~(1982), ``An approach to time series smoothing and forecasting using the EM algorithm,''  {\it Journal of Time Series Analysis}, Vol.\ 3, pp.\ 253--264.

 \bibitem[Spearman (1904)]{Spearman04}C.\ Spearman~(1904), ``{\it General intelligence}, objectively determined and measured,''  {\it The American Journal of Psychology }, Vol.\ 15, pp.\ 201--292 .
 
 \bibitem[Stock and Watson (2002a)]{SW02a}J.H.\ Stock and M.W.\ Watson (2002a), ``Forecasting using principal components from a large number of predictors,''  {\it Journal of the American Statistical Association}, Vol.\  97, pp.\ 1167--1179.
  
  \bibitem[Stock and Watson (2002b)]{SW02b}J.H.\ Stock and M.W.\ Watson (2002b), ``Macroeconomic forecasting using diffusion indexes,''  {\it Journal of Business \& Economic Statistics}, Vol.\  20, pp.\ 147--162.
  
  \bibitem[Stock and Watson (2016)]{SW16} J.H.\ Stock and M.W.\ Watson~(2016), ``Dynamic factor models, factor-augmented vector autoregressions, and structural vector autoregressions in macroeconomics,''  in: J.B.\ Taylor and H.\ Uhlig, Eds., {\it Handbook of Macroeconomics}, Vol.\ 2, pp.\ 415–525, Elsevier.
  
   \bibitem[Tang et al.~(2021)]{} C.~Tang, H.L.~Shang,  and Y.~Yang (2021), ``Multi-population mortality forecasting using high-dimensional
functional factor models,''  arXiv:2109.04146.
  

   \bibitem[Tavakoli et al. (2023)]{Tava23}S.\ Tavakoli, G.\ Nisol, and M.\ Hallin~(2023), ``Factor models for high-dimensional
functional time series
II: Estimation and forecasting,''  {\it Journal of Time Series Analysis}, Vol.\ 44, pp.\  601--621. 

 \bibitem[Tiao and Tsay (1989)]{TT89} G.C.\ Tiao and R.S. Tsay~(1989), ``Model specification in multivariate time series,''  {\it Journal of the Royal Statistical Society: Series B (Methodological)}, Vol.\ 51, pp.\ 157--195.
 
 \bibitem[Tipping and Bishop (1999)]{TB99}M.E.\ Tipping and C.M.\ Bishop~(1999), ``Probabilistic principal component analysis,''  {\it Journal of the Royal Statistical Society: Series B (Statistical Methodology)}, Vol.\ 61, pp.\ 611--622.

 \bibitem[Trapani (2018)]{Trap18} L.\ Trapani (2018),  ``A randomized sequential procedure to determine the number of factors,''  {\it Journal of the American Statistical Association}, Vol.\ 113, pp.\ 1341--1349.
 
   \bibitem[Truc\'{\i}os et al.~(2019)]{} C.~Truc??os, L.K.~Hotta, and P.L.~Valls Pereira~(2019), ``On the robustness of the principal volatility components,''  {\it Journal of Empirical Finance}, Vol.~52, pp.~201--219.

 
   \bibitem[Truc\'{\i}os et al.~(2021)]{}  C.~Truc\'{\i}os, J.H.G.\ Mazzeu,  L.K.\ Hotta,   P.L.~Valls Pereira, and M. Hallin~(2021), ``On the robustness of the general dynamic factor model with infinite-dimensional space: identification, estimation and forecasting,'' {\it International Journal of Forecasting}, Vol.~34, pp.~1520--534.
   
 
 
   \bibitem[Truc\'{\i}os et al.~(2022)]{Trucetal22} C.~Truc\'{\i}os,  J.H.G.\ Mazzeu, M.\ Hallin,  M.\ Zevallos, L.K.\ Hotta,  P.L.\ Valls Pereira~(2022), ``Forecasting conditional covariance matrices in high-dimensional time series with application to dynamic portfolio optimization: a General Dynamic Factor approach,''    {\it Journal of Business \& Economic Statistics}, Vol.\ 41, pp.\  40--52.
   
   \bibitem[Yamamoto and Tanaka (2015)]{}Y.\ Yamamoto and S.\ Tanaka~(2015), ``Testing for factor loading structural change under common breaks,''  {\it Journal of Econometrics}, Vol.\ 189, pp.\ 187--206.


   


%
%
%
%
%
%

\end{thebibliography}
\end{document}